# Duality of the collective and single particle responses in simple metals in the extreme long wavelength limit.

K. Kempa, *Department of Physics, Boston College.*

## Abstract

It is demonstrated that the collective and single particle responses of simple metals in the extreme long wavelength limit become identical/dual. When applied to alterative models of a metallic slab, this duality proves equivalence of the *plasma* frequency (in units of energy) and the surface *energy barrier*, which confines electrons inside the metal. This in-turn proves also a simple, yet powerful formula, which expresses the work function of a metals as a difference between its free-electron plasmon and Fermi energies, and which has been shown to be more accurate (even for complex metals) than the best available "ab-initio" simulations.

In a seminal paper [1], Walter Kohn showed that the electromagnetic response in the extreme long wavelength limit (ELWL), of an interacting electron gas subject to a constant magnetic field is, in general, identical to that of a single electron in this confinement. This so called Kohn's Theorem (KT) has been later reformulated for an interacting electron gas confined by a parabolic potential in wide parabolic quantum wells [2], and semiconductor quantum dots [3]. Here we show that this duality between single-particle and collective responses, implies equality of the *plasma* frequency (in units of energy) and the surface *energy barrier*, which in an effective single electron picture confines electrons inside the metal. This immediately proves a simple formula, which expresses the work function of a metals as a difference between its free-electron plasmon and Fermi energies, and which has been shown [4] to be more accurate (even for complex metals) than the best available "ab-initio" simulations.

Consider a thick slab of a "jellium" metal, which consists of *N* electrons (electron charge *e* and mass *m*) with average charge density *n*, confined by a uniform positive charge (jellium) of macroscopic thickness *W*, and charge density $n_+ = n$, as sketched in Fig. 1(a). The electrostatic confining potential produced by this slab is parabolic

$$V(z) = \frac{1}{2} m\omega_0^2 z^2 \qquad \text{for } |z| \le W/2 \qquad (1)$$

and linear

$$V(z) = \frac{1}{2} m\omega_0^2 W\left(|z| - \frac{W}{4}\right) \qquad \text{for } |z| > W/2 \quad , \qquad (2)$$

with

$$\omega_0 = \sqrt{4\pi n_+ e^2 / m} = \omega_p \ , \qquad (3)$$

where $\omega_p$ is the plasma frequency of the electron gas.

This potential is sketched in Fig. 1(b).

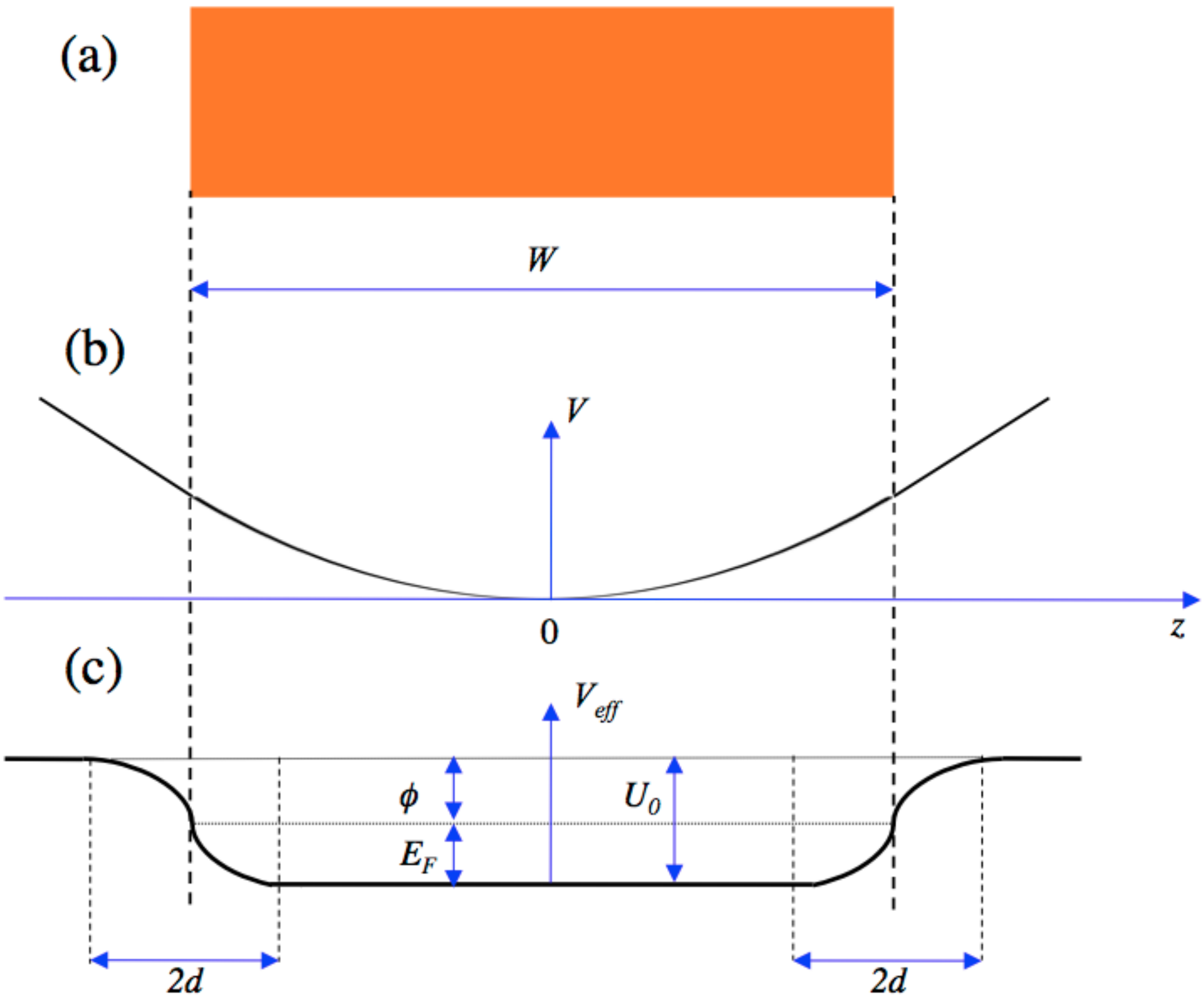

**Fig. 1.** (a) Schematic of a slab of positive charge ("jellium"); thickness of the slab is W, and it is extended in the *x-y* plane. (b) The electrostatic potential produced by the slab. (c) The effective single-electron potential in the self-consistent calculation.

The vast majority of electrons in the slab see the parabolic potential, given by Eq. (1), since the "spillage" of the electrons outside the "jellium" is very small, of the order of a few angstroms, while $W$ is assumed macroscopic. Thus, one can immediately apply KT, and show that in ELWL response of the slab to the electromagnetic wave (polarized perpendicular to the slab) is identical to that of a single electron in this potential, i.e. the absorption occurs for

$$\omega = \omega_0 = \omega_p \tag{4}$$

To demonstrate this explicitly consider the Hamiltonian of the *N*-electrons in the slab

$$H = \sum_{i=1}^{N}\left[\frac{\mathbf{p}_i^2}{2m} + V(z_i)\right] + \frac{1}{2}\sum_{i=1}^{N}\sum_{j=1}^{N} u(r_{ij})$$

For *V* given by Eq. (1) this equation takes a form

$$H = H_{cm}(\mathbf{P}, \mathrm{Z}) + \frac{1}{2N}\sum_{i=1}^{N}\sum_{j=1}^{N} H_{rel}(\mathbf{p}_{ij}, r_{ij}, z_{ij}) \tag{5}$$

where the center of mass part, which depends *only* on the center of mass variables $\mathbf{P} = \sum_{i=1}^{N}\mathbf{p}_i$ and $Z = \sum_{i=1}^{N} z_i / N$ is given by

$$H_{cm} = \frac{\mathbf{P}^2}{2mN} + \frac{1}{2} mN\omega_0^2 Z^2 \tag{6}$$

A set of independent on **P** and Z variables $\{\mathbf{p}_{ij} = \mathbf{p}_i - \mathbf{p}_j, r_{ij} = r_i - r_j, z_{ij} = z_i - z_j\}$ enters *only* the second part of the Hamiltonian in Eq. (5). The radiation dipole operator for the *z*-polarized radiation in ELWL, given by $D_z = e\sum_{i=1}^{N} z_i = eNZ$, is a function of *only* the center of mass variables, and therefore couples *only* to $H_{\mathrm{cm}}$. Thus, regardless of *N*, absorption in this case occurs at a single-particle frequency $\omega_0$, and causes a simple, collective and synchronous harmonic motion of all *N* electrons. This is an exact result, and it demonstrates the *duality of the single-particle and collective motion/response* in this case. This is a general property of all systems satisfying KT.

One of the consequences of the duality is that it allows the following, purely single-particle interpretation of the electron excitations: they remain single-particle in the presence of an arbitrary number of electrons. This interpretation allows a constructive application of the duality, and we apply it now to derive a simple expression for the work function of metals. To accomplish that, we consider now the well-known alternative description of the slab system [5,6]. In this mean-field approach, each electron "sees" an effective potential $V_{eff}[n;z]$, which depends explicitly only on $z$ and the electron density $n$, but *not* the electron energy $E$. This potential is a sum of the confinement potential (Eqs. 1 and 2), the Hartree potential of all other electrons, and the exchange and correlation potential. The corresponding eigenvalue problem takes a one-dimensional, single-electron form [5]

$$\left\{-\frac{\hbar^2}{2m}\frac{d^2}{dz^2}+V_{eff}[n;z]\right\}\psi_E(z)=E\psi_E(z) \tag{7}$$

where

$$n=\frac{1}{\pi}\int_0^{k_F}(k_F^2-k^2)\left|\psi_E(z)\right|^2dk \tag{8}$$

$\hbar$ is the Planck constant, and $k_F$ is the Fermi wave vector. The coupled eqs. (7) and (8) form a self-consistency loop, and can be solved only numerically, and only for a modeled exchange and correlation part of $V_{eff}[n;z]$ [5]. For the purpose of this work we do not need details of the solutions, since except in a very narrow surface region of thickness $d$ << W, $V_{eff}$ is constant: zero well inside the slab, and $U_0$ well outside as shown in Fig. 1(c). In those flat/constant regions Eq. (7) has simple solutions. Since the states are extended, the dispersions from each flat region apply to states in the whole space, and thus form two bands. The first band has dispersion $E=\hbar^2k^2/2m$ (originating from the region $|z| \leq W/2 - d$), and the second band $E' = E + U_0$ (originating from the region $|z| > W/2 + d$).

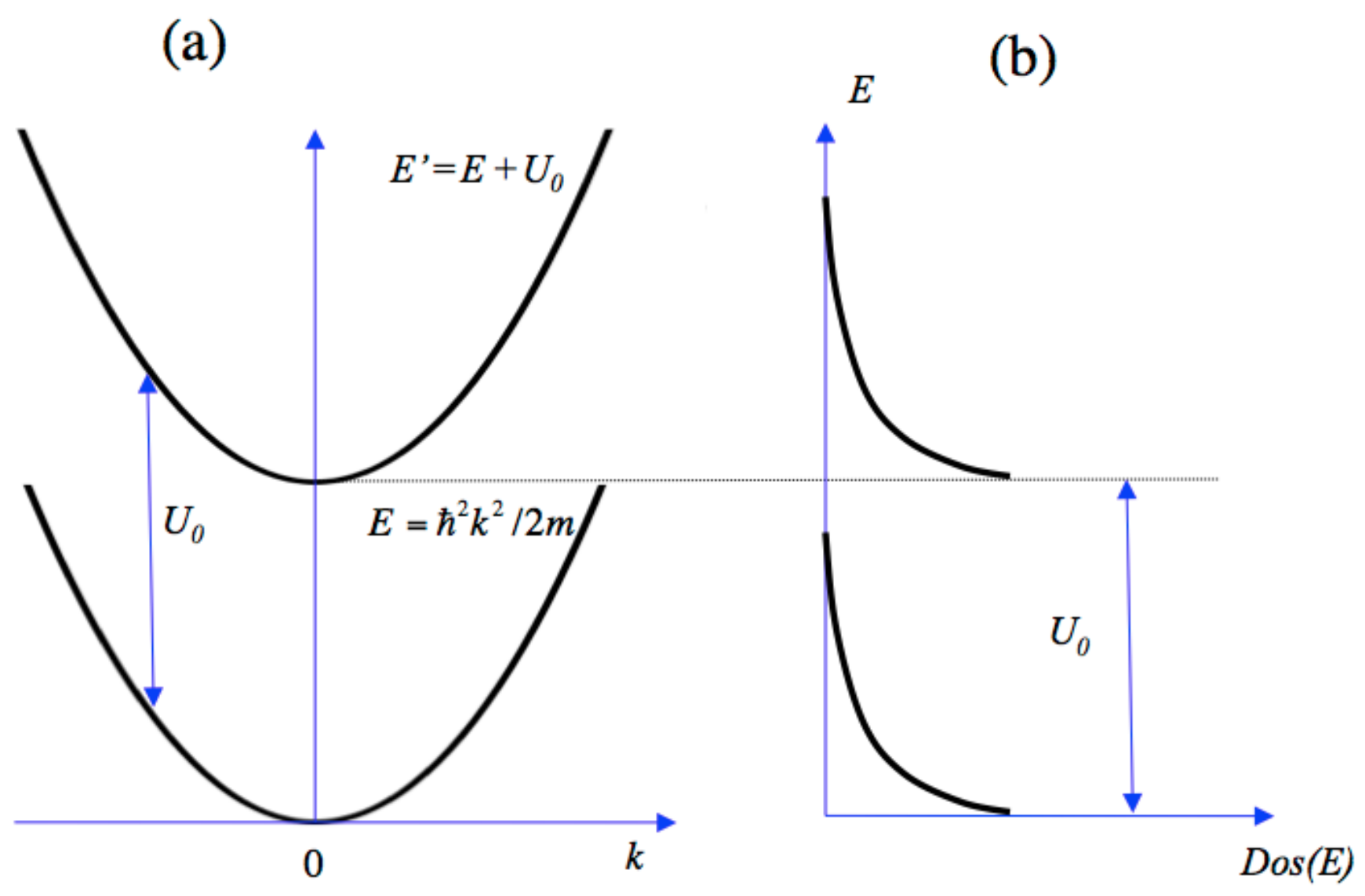

**Fig. 2.** (a) The single-electron dispersions in the mean-field self-consistent calculation, and the corresponding density of states (b).

Since the duality of the collective and single-particle excitations must hold, as demonstrated in the KT approach, so must the possible interpretation of the electron excitations as single-particle only. Accordingly, in the present mean-field approach we can calculate only the single-electron absorption (in ELWL, and of a *z*-polarized radiation of frequency $\omega$), for example from the Fermi's Golden rule [7]

$$\alpha(E,\omega) = A\omega\langle z\rangle^2 \Omega(E,\omega), \tag{9}$$

with $A = 4\pi^2 e^2/\hbar$, and

$$\langle z\rangle = \int_{-\infty}^{\infty} \psi_{E'}^*(z) z \psi_E(z) dz \tag{10}$$

where $E' = E + \hbar\omega$. $\Omega(E,\omega)d\omega$ is the number of states having vertical ($q$ = 0) energy difference $\hbar\omega$ between the bands in the range $d\omega$. It is clear from Fig. 2 (a) that the *only* allowed vertical energy difference is $U_0$. Consequently, $\Omega(E,\omega)$ and therefore also the absorption coefficient $\alpha$ is nonzero (i.e. the ELWL absorption occurs) *only* for $\hbar\omega = U_0$. This result must be obviously consistent with the corresponding KT result, and thus finally, according to Eq. (4) and Fig. 1(c)

$$\hbar\omega_p = U_0 = E_F + \phi \qquad (11)$$

This simple relation between the work function $\phi$ of a simple metal, the plasma frequency $\omega_p$ and the Fermi energy $E_F$, has been known in the past, but until now never derived. Here it is shown to be a natural consequence of the essentially parabolic electronic confinement in the metallic slab, and the resulting duality of the single-particle and collective response in ELWL.

Eq. (11) provides an explicit expression for the work function in terms of $n$

$$\phi = \left(\hbar\sqrt{\frac{4\pi e^2}{m}}\right)n^{1/2} - \left[\frac{\hbar^2}{2m}\left(3\pi^2\right)^{2/3}\right]n^{2/3} \qquad (12)$$

Recently it was demonstrated [4], that remarkably, this formula agrees with the experiment better than *any* currently available numerical simulation. Moreover, this equation, which uses only free-electron parameters, works well not only for simple (s and p) metals, but also for more complex metals ($s^2$, $d^{10}s^2$, $s^2p^2$, and $s^2p^3$). This unusual "robustness" of Eq. (11) has a few origins. Firstly, any atomic scale corrugation of the positive ionic charge (above assumed to be uniform) will be averaged out by the extremely long wavelength field, and thus its importance suppressed. Secondly, in the mean-field treatment, we note that the orthogonality of the eigenstates implies that $\langle z\rangle \neq 0$ for transitions between bands, and therefore this term is not interfering with the derivation of Eq. (11). Also, the full expression for $\Omega(\omega)$ in Eq. (9) can be obtained by first noticing, that both bands have the same density of states (*Dos*), offset by the energy shift $U_0$

$$Dos(E) = \frac{L}{2\pi}\left(\frac{dE}{dk}\right)^{-1} = \frac{B\theta(E)}{\sqrt{E}} \qquad (13)$$

where $B = L\sqrt{m}/2\pi\hbar\sqrt{2}$. Sketch of these densities of states is shown in Fig. 2(b), demonstrating that these are singular at $E = 0$, and $E = U_0$. Therefore

$$\Omega(\omega) = Dos(E)\delta_{\hbar\omega,U_0} = B\theta(E)\delta_{\hbar\omega,U_0}/\sqrt{E}\,, \qquad (14)$$

which shows that the dominant absorption occurs for transitions between the lowest energy states of the respective bands. Thus, the derivation of Eq. (11) remains valid even if the corresponding bands are not identical. The derivation will brake-down only for metals with two, or more distinct groups of electron states in the valence band, such as is the case for the complex noble metals like Au and Ag ($d^{10}s^1$). Indeed, in that case while Eq. (11) predicts $\phi_{duality} \approx 3.5eV$ (for both metals), the measured values are $\phi_{Ag} \approx 4.3eV$ and $\phi_{Ag} \approx 5.1eV$. Since this discrepancy seems not too severe (20-30% error) and understandable, Eq. (11) could be modified/extended to be valid even in such complex systems.

In Conclusion, by employing the Kohn's theorem it was demonstrated that the collective and single particle responses of simple metals in the extreme long wavelength limit become identical/dual. Consequently, the single-particle picture can be used: excitations of electrons can be viewed as single-particle regardless of the presence of other electrons. When applied to an alterative model of the metallic slab, this proves equivalence of the *plasma* frequency (in units of energy) and the surface *energy barrier*, which confines electrons inside the metal. This in-turn proves also a simple formula, which expresses the work function of a metals as a difference between its free-electron plasmon and Fermi energies, and which has been shown to be more accurate (even for complex metals) than the best available “ab-initio” simulations.

**Acknowledgements**

The author is grateful to Prof. Frank Forstmann for bringing this problem to his attention, and to Prof. Andrzej Herczynski for careful reading of the manuscript, and useful comments.

## References

[1] W. Kohn, *Phys. Rev*. **123**, 1242 (1961).

[2] L. Brey, N. F. Johnson, and B. I. Halperin, Phys. Rev. B 40, 10647 (1989).

[3] P. Bakshi, D.A. Broido, and K. Kempa, *Phys. Rev.* B **42**, 7416 (1990).

[4] F. A. Gutierrez, J. Diaz-Valdes, H. Jouin, *J. Phys.:Condens. Matter* **19** (2007) 326331.

[5] N.D. Lang and W. Kohn, *Phys. Rev.* B **1**,4555(1970).

[6] N.D. Lang and W. Kohn, *Phys. Rev.* B **3**,1215(1971).

[7] J.M. Ziman, "*Principles of the theory of solids*", Cambridge University Press, second edition, 1972.